# Advanced Biophotonics techniques for cell and molecules manipulation associated with cancer and autoimmune diseases – the role of optical tweezers

Ellas Spyratou[1,2*]

[1] 2nd Department of Radiology, Medical School, National and Kapodistrian University of Athens, 11517 Athens, Greece
[2] School of Medicine, Democritus University of Thrace, Alexandroupolis, Greece

*__Corresponding author__
Spyratou Ellas
Email: ellas5@central.ntua.gr

Biophotonic techniques are growing in rapid rhythms enabling the monitoring of subcellular structures and non-invasive theranostic interventions in cancer and autoimmune diseases. The integration of Biophotonics with nanotechnology and biosensors brings a revolution in the micro- and nano-world with new optical tools. Among them, optical tweezers revive as a potential tool for tracking cells behavior and probing interactions forces between cells, cells-biomolecules and cells-nanoparticles. In this review we aim to exhibit the state-of the art advances of the Biophotonics in the diagnostic and therapeutic field, and the role of optical tweezers.

**Introduction**

Rapid advances in Biophotonics are revolutionizing the illumination of several diseases and, among them, cancer pathogenesis and autoimmune disease processes. Biophotonics is an emerging multidisciplinary research area, embracing all light-based technologies applied to the life sciences and medicine. The expression itself is the combination of the Greek syllables' "bios" standing for life and "phos" standing for light [1]. In Biophotonics, the "conventional" light is monochromatic laser or laser-like non-ionizing radiation and the basic biomedical applications to all levels of biological structures are divided into two major fields. The first is devoted to diagnostic and imaging applications (*in vivo* and *in vitro*, in cellular or molecular level) and the second is therapy or surgery by radiation (e.g. biostimulation, tissue removal-surgery, photodynamic therapy, cell micromanipulation) [2]. Modern non-invasive laser-based optical research techniques prove to be more and more useful in the biomedical field, covering very different domains, from practical, clinical applications to molecular and cellular biology fundamental research. Today, several efforts aim to miniaturization of Biophotonics tools, leading to a highly researched field, namely Nanobiophotonics, which refers to the research and development of novel technologies, biosensors, and drug delivery systems for prevention, diagnosis, and treatment of diseases at the nanoscale, in sub-cellular and molecular level [3] and for the dream of personalized therapy.

Recent advantages in cancer therapy couples Biophotonics with nanoparticles for a more effective and targeted treatment. Metallic nanoparticles, semiconductor quantum dots and carbon nanotubes have been used as photosensitizer agents for photo-triggered diagnosis and photo-triggered therapy [4, 5]. Nanocarriers can be served as multi-modal theranostics systems [6] or as "Trojan Horses" [7] carrying multiple therapeutic and imaging agents directed to tumors. They can provide simultaneously diagnostic information and targeted drug delivery by photon-based stimulus through fluorescence, photothermal and photodynamic treatment [8].

In recent decades, biomedical research has contributed significantly to understanding autoimmune diseases. Researchers are conducting laboratory and clinical studies comparing various parameters of the immune system. An active field of study is the identification of genes involved in genetic abnormalities during cell apoptosis. However, the cause of autoimmune diseases is still a question that has not been answered worldwide and may be due to a combination of genetic, environmental and hormonal factors. In immunology, many cellular functions associated with the transduction of cellular signals by lymphocytes have been extensively studied by biochemical methods [9]. In some cases, such as in the action of T cells, natural contact between an antigen and a lymphocyte is necessary for the antigen receptor to recognize and bind to it. Applied imaging techniques such as FRET (fluorescent resonance energy transfer), FRAP (fluorescence recovery after photobleaching) and SPT (single-particle tracking) methods made it possible to visualize the signal between lymphocytes with great spatial and temporal resolution at the cellular and molecular level [10].

Among the plethora of photon-based techniques, optical trapping, with the ability of applied Biophotonics interventions in living cells, is a very promising tool for improving the understanding of cancer and the autoimmune mechanisms of cells [11, 12]. Optical tweezers technique is a non-invasive biomedical tool with advanced applications in biology, genetics, medicine, and nanotechnology. The ability to "touch" the microcosmos non-invasively, while performing nanometer-precision and submicrometric analysis, using a single optical tool, is a revolutionary technique. Optical tweezers can manipulate cells, viruses, bacteria and macromolecules. By using the optical trapping technique, a cell can be selectively, non-invasively and non-destructively manipulated to a phagocyte, to attach to its surface cell receptor and trigger the initiation of the phagocytosis process [13]. Optical tweezers are capable to grab, tracking and manipulate small virus such as influenza and SARS-CoV-2 or even the protein corona-decorated entities [14].

In this review, we briefly highlight the novel photon-based theranostics modalities for human diseases, remembering at this point that several aspects of cancer confrontation have similarities with autoimmune diseases pathogenesis and treatment.

## 2. Biophotonics and theranostics

### 2.1. Biophotonics in diagnosis

As we mentioned, Biophotonics is a relatively novel interdisciplinary discipline that integrates lasers, optoelectronics, photonics and biomedical sciences, dealing with the interaction between non-ionizing light quanta and biological materials, including tissues, cells and even sub-cellular structures and molecules in living organisms [15]. In theranostics applications of Biophotonics, the light-based techniques for monitoring and treating diseases permit minimally invasive interventions that reduce patient discomfort [16]. Like the conventional radiography, ultrasound (US), and magnetic resonance imaging (MRI) diagnostic modalities, most of the non-ionizing radiation launched imaging techniques were first developed for cancer detection and rapidly were applied also to several diseases, including autoimmune disorders [17,18].

A careful search in the literature reveals a variety of research and clinical studies based on cellular and sub-cellular diagnosis *via* flow cytometry, light-microscopy techniques, and laser-induced fluorescence spectroscopy. To completeness, we just refer here the titles of a few imaging techniques, as: Epifluorescence microscopy, immunofluorescence microscopy, confocal microscopy, Total Internal Reflection Fluorescence - TIRF microscopy, Two-Photon Laser Scanning Microscopy – TPLSM, Fluorescent Resonance Energy Transfer – FRET, PhotoActivated Localization Microscopy - PALM [10]. Figure 1 shows an immunofluorescence microscopy image of lymphocytes immunostaining with primary γ-H2AX antibody, secondary with Rhodamine Red-X anti-rabbit fluorescent antibody and finally with nuclear stain DAPI for the detection of DNA double-strand breaks [19, 20]. A decade ago, Balagopalan *et al* reviewed the application of imaging techniques to the study of lymphocyte activation, starting the hierarchy of imaging scale from the whole animal or tissue level and further zooming in to the cellular and subcellular levels, at molecular resolution [10].

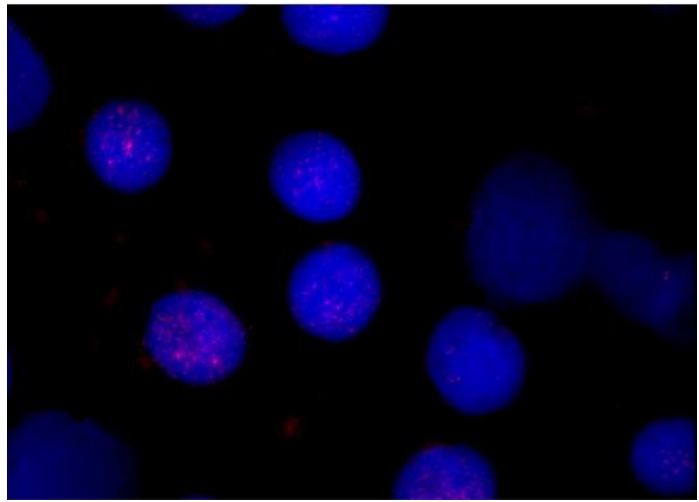

Figure 1. Immufluorescence image of lymphocytes stained for γ-H2AX detection. Red spots correspond to γ-H2AX foci which are strong correlated with DSBs; blue, nucleus stained with DAPI.

Multiple imaging modes can offer complementary information and overcome the limitations of each single modality. The combination of optical imaging with computed tomography (CT), magnetic resonance imaging (MRI), Positron Emission Tomography (PET) and Single-photon emission computed tomography (SPECT) can enhance size resolution and penetration depth [21, 22]. Near-Infrared Fluorescence (NIRF) imaging is highly attractive for early non-invasive detection of cancer due to its high penetration depth and low autofluorescence. Different types of multicomponent nanoparticles like PEGylated Au/SiO2 nanocomposites conjugated with Fluorescein isothiocyanate (FITC) [23], Iron Oxide NPs encapsulating in Human Serum Albumin (HAS) [24] etc have been designed to act as dual contrast agents offering multimodal imaging (Figure 2). The combination of CT, MR, PET and SPECT modalities with fluorescence imaging modalities can allow extension of imaging across the dynamic range of size resolution and penetration depth from deep-body with size resolution of ~1mm to thin penetration depths of few hundred microns or millimeters with size resolutions to single-cell or even subcellular resolution.

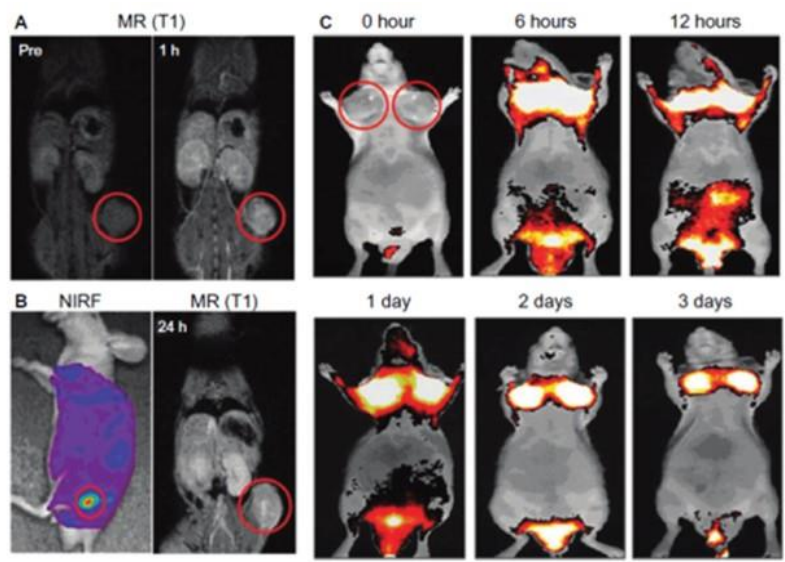

Figure 2. (A-C) *In vivo* magnetic resonance (MR)–near-infrared fluorescent (NIRF) dual-modality imaging of SCC7-bearing mice. Cy5.5-chitosan nanoparticle-Gd(III) nanoparticles were injected into the SCC7-bearing mice, and the mice were visualized by using MR and NIRF imaging. Red circles indicate tumor sites. (A) *In vivo* MR imaging showed T1-positive contrast effects 1 hour after injection at the tumor sites. (B) Both NIRF and T1-weighted MR images were simultaneously observed after 1 day post injection of Cy5.5-CNP-Gd(III). *In vivo* NIRF imaging showed brighter NIRF intensity at the tumor site. In vivo MR imaging 24 hours after injection showed bright contrast effects at the tumor site. (C) *In vivo* NIRF imaging showed the accumulated Cy5.5-CNP-Gd(III) nanoparticles over time at the tumor sites. Reprinted (adapted) with permission from ([25], Nam et al., 2010). Copyright (2021) American Chemical Society.

Recently, the introduction of novel diagnosis modalities is tested, and namely the role of biosensors specialized for sensing enzymes, cells, and DNA, RNA, or antibodies. A biosensor is an analytical tool sensitive to a biochemical stimulus, which converts a biological response into an electrical or optical signal by transmitting information about a vital process. In 2020, Imas *et al* published a comprehensive review on optical biosensors for the detection of biomarkers associated with RA (rheumatoid arthritis), including microRNAs (miRNAs), C-reactive protein (CRP), rheumatoid factor (RF), anti-citrullinated protein antibodies (ACPA), interleukin-6 (IL-6) and histidine, which are biomarkers that enable RA detection and/or monitoring [26]. Certainly, the clinical translation of biosensors for RA diagnosis requires a clear connection between the selected biomarkers and RA, with high specificity and sensitivity exclusively, as possible, for the corresponding disease.

## 2.2. Biophotonics in therapeutics

Recent theranostics techniques are combined with nano-imaging and nanomaterial-based drug delivery techniques for an effective and targeted disease management. In modern anti-cancer modalities, the majority approaches of delivering new drugs are behaving as a "Trojan horse", by introducing the active, cytotoxic compound in a nanoparticle, "decorating" sometimes its surface with a ligand that trigger the cancer cell into taking it up. Imaging can be used to trace the delivery of the drug inside the body and simultaneously to activate the release of the drug by an external stimulus such as laser light. Functionalized nanoparticles can act both as contrast agents and photosensitizers for photothermal (PT) [27] or photodynamic treatment (PDT) [28]. Among the large variety of NPs, metal NPs (AuNPs) are in the cutting edge of the nanomedicine due to their unique physical, optical and electronic properties [29, 30]. When metal NPs excited by visible or infrared monochromatic light with laser wavelength corresponding to their Surface Plasmon Resonance (SPR), the conduction electrons of the metal can be subjected to coherently oscillation and convert the electromagnetic energy into heat providing targeted tumour disruption via hyperthermic damage [31]. Photodynamic therapy has come again to the forefront due to the new class of photosensitizers (PS) which enhance PTD efficiency. The encapsulation of PS such as verteporfin or methylene blue into nanocarriers seems to overcome some of the barriers of the PS i.e poor selectivity to the target tissues, the low extinction coefficients, their lipophilicity, the photobleaching of the PS e.t.c. The synergia of nanomedicine with biophotonic techniques could lead to a localized "surgery" causing tumor disruption or removal without invasiveness [28].

Nano-image guided surgery plays an emerging role in the field of personalized tumour surgery. Fluorescence-imaging guided surgery can be used for sentinel lymph node mapping or to distinguish the margins of a tumor in microscopic scale and in real time. Nanoparticles like quantum dots, liposomes or supermagnetic NPs can be conjugated with fluorescence dyes acting as targeted optical imaging probes to offer high selectivity and specificity [32].

Drug loaded nanocarriers like magnetic NPs or gold NPs have been surface functionalized with monoclonal antibodies (mAbs) acting as optical molecular image probes and at the same time as therapeutic agents [33-35]. Therapeutic mAbs are in great interest as drugs against diseases such as autoimmune disorders, inflammatory diseases and cancer due to their unique properties [36, 37]. They demonstrate high specificity and affinity to cell antigens Thus, monoclonal antibodies enhance the stability of the nanosystems and prolong their half-time circulation.

### 3. The role of optical tweezers in Biophotonics

Over 30 years of exploration after the first report of damage-free optical trapping of virus and bacteria, [38] optical tweezers have found innumerable applications in cell biology and living systems studies. The major contribution of optical trapping and manipulation to the biological sciences gained acceptance and worldwide recognition in 2018, with the award of half the Nobel Prize in Physics to pioneer Arthur Ashkin, the "father" of optical trapping, who passed away in September 2020, at the age of 98. The optical trapping technique uses one or more laser beams to selectively manipulate position, motion, and dynamics of micro- and nanostructures. This phenomenon is based on the optical forces of the order of some piconewtons exerted by laser radiation when it interacts with matter. The optical force exerted on a cell depends on its surface and the cytoplasmic refractive index. Therefore, the biochemical changes that happens in the cell cytoplasm or membrane reflects its behavior under the optical trap [39].

The contribution of optical trapping technique is particularly important in the field of biomedical science, as it is possible to study the properties of cells at a primary level, overcoming the problems of conventional techniques that examine cell populations. Published studies report the use of optical traps to selectively bring killer cells into contact with other target cells. The response of their immune system was studied, while it was possible to visually observe the change in the morphology of a defenseless target cell, as it comes in contact with a killer cell [40-42].

Optical tweezers are used extensively for studying living cells, e.g for hemorheology studies, blood microcirculation and biomechanical properties [43, 44]. Particularly important is the study of the elastic properties of erythrocytes with the help of optical trapping. The researchers used optical tweezers to elongate human erythrocytes through the dual optical trapping of silicone spheres attached to the cell membrane or using line optical tweezers (Figure 3) and were able to determine their degree of torsion. The shear modulus was calculated by measuring the cell membrane deformation in function with the optical forces exerted to the membrane *via* small optically trapped silica beads [45]. The bending modulus of the membrane was estimated by measuring erythrocyte's folding time in function with laser power under the effect of line optical tweezers [46]. Optical tweezers have been used to induce rotation and folding of erythrocytes to study their elastic properties and diagnose malaria in them [47-48]. The electrical and mechanical properties of erythrocytes have recently been measured using a dual trap. The Z-potential of erythrocytes, which prevents them from aggregation, was calculated by measuring the rate at which a trapped erythrocyte escaped from the trap into an electrolyte solution [49].

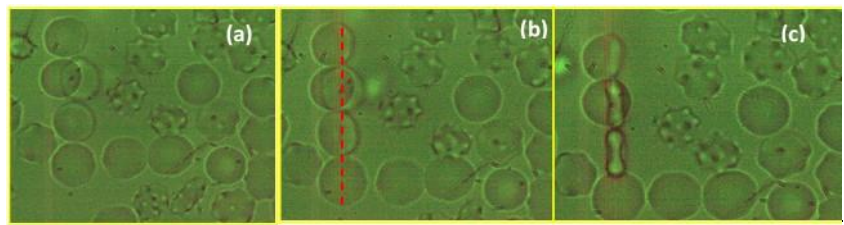

Figure 3. (a) Three red blood cells under a line optical tweezers. The RBCs are trapped simultaneously, (b) are folded gradually and (c) orient its long axis in the direction of the electric field of incident beam. - - - Dotted red line is the direction of the line optical trap.

In addition to other biomedical areas, there have been great advances in optical trapping and its combination with other Biophotonics tools in neuroscience research, for studying the physical properties and intrinsic forces of neurons, their communication modalities, as well as some of the fundamental neuronal growth and dynamics function [50]. In their comprehensive review, Lenton *et al* ascribe the long time from the first development of optical tweezers in biology (1987) to specific applications in neuroscience, partially to the complexity of the brain and the associated difficulties with trapping or imaging objects within it [50]. In parallel, last year Zhao *et al* [51] proposed and implemented an optical shield scheme, based in far-field Bessel beam, for manipulating individual cells in a crowded environment (e.g. single blood cell, individual lymphocytes from an inguinal lymph node).

So far, no clinical application in human has been implemented by using optical tweezers. Infrared optical tweezers have been used to trap and manipulate erythrocytes in the blood capillaries in the ear of a mouse. Optical tweezers were capable to interfere to the blood stream by trapping erythrocytes or removing a blockage [52]. The ability to selectively manipulate single cells can have many advantages *in vivo* for the diagnosis and treatment of metastatic cancer cells which can travel through the bloodstream and the lymph system. Moreover, understanding the biophysics of individual cell deformation offers the means for new perspectives in prognosis, disease diagnosis, and treatment. Changes in the ability to deform cell shape, combined with changes in cell adhesion affect cell reproduction, cell signal transmission, and cell metastasis potential [53]. The development of various multifactorial diseases, such as systemic lupus erythematosus - a chronic autoimmune disease - is impaired by pathologically enhanced erythrocyte accumulation (Red Blood Cells, RBCs) [54]. A particular increase in RBCs accumulation forces is a clinical symptom of the disease. The elastic properties of individual RBCs determine their ability to resist large deformations by passing through small capillaries. Therefore, the study of their biomechanical properties is of particular importance for the study of immune-regulatory mechanisms. Optical tweezers can be used as passive "force clamps" to induce and study elastic deformations in individual cells [55]. They are a tool for calculating the stiffness and torsion rate by measuring sub-micrometric cell deformations, which are caused by optical forces. Optical tweezers can be combined with Raman spectroscopy to characterize and monitor the physical and chemical properties of cells. For example, Laser tweezers Raman spectroscopy was used to monitor the changes in the oxygenation state of human red blood cells while they were stretched with optical tweezers [56, 57].

High-precision optical tweezers have been developed for protein folding experiments. Optical tweezers were used to apply mechanical forces and to monitor proteins unfolding. Recent studies correlate the folding and misfolding of human membrane proteins in autoimmune diseases [58]. The way that proteins fold from linear chains to three-dimensional structures or *vice versa* is under great interest in biology. Single proteins or protein complex can be tethered between two microbeads by using DNA linkers or antibody linkers. The beads are trapped by using a dual optical tweezers and can be pulled away by alter the distance between the laser traps. Thus, protein unfolding is induced by the mechanical forces and the kinetics of the protein can be monitoring [59, 60]. The mechanical forces at which folding transitions take place depend on the pulling speed [60].

A very challenging demand is the miniaturization of the biophotonic set-ups opening new, fascinating possibilities for *in vitro* single cell experiments like flow cytometry, laser induced fluorescence and for *in vivo* detection of disease. Optical tweezers can be coupled with nanophotonic biosensor devices based on integrated fiber optics and microfluidics devices for the implementation of lab-on-a-chip platforms. They can probe complex biophysical and biomechanical processes governing cell-cell interactions, cell – surface interactions, cell sorting and drug delivery/testing [61-63]. Figure 4 illustrates the implementation of advanced optical trapping techniques (a) for probing erythrocytes biochemical or biomechanical properties and (b) integrating with microfluidics devices for performing single cell manipulation and sorting between target with NPs cancer cells and healthy cells. Moreover, the selective observation of cells-nanoparticles interactions will lead to a better understanding of the interaction mechanisms and to more targeted treatments.

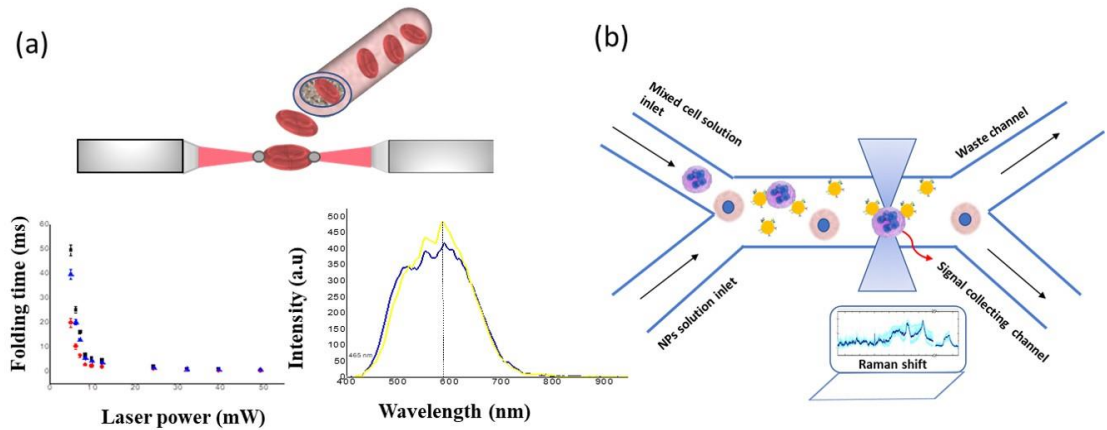

Figure 4. Schematic representation of advanced optical trapping applications (a) Optical tweezers act as a tool for the evaluation of erythrocyte's deformability which is an important biomarker for circulation efficiency and the probing of any biochemical modification of membrane proteins associated with cell health (b). Dual-beam optical tweezers coupled with microfluidics device to manipulate and recognize targeted with functionalized NPs tumour cells from healthy cells. In both cases, the systems can be integrated other spectroscopic set-ups such as dark-field microscope and Raman microscope collecting signals like the spectra from RBC membrane features. Moreover, from the images acquired by the CCD camera, RBC deformations can be measured.

**4. Concluding remarks and future perspectives**

From the beginning of 21$^{st}$ century to nowadays, OTs gain great ground in the field of Biophotonics. As it was already mentioned, optical trapping works by combining radiation pressure phenomena for developing optical forces. Historically, the idea that light transmits momentum and therefore can exert forces on electrically neutral objects has been attributed to Kepler and Newton and was the basis for comets' tail explanation. The novelty is, in our days, that we can beneficiate from Physics achievements to move from Astrophysics to Nanomedicine, through a wonderful journey into the paradoxes of the quantum world. In this enlightened journey, let us make two interesting stops:
(a) Recently, upconverting nanoparticles (UCNPs) have been introduced in Biophotonics and nanophotonics as very promising theranostic intratissue probes in biological tissues [64, 65]. Their unique property to convert near-infrared (NIR) light into visible or ultraviolet light *via* photon upconversion mechanism will permit interventions to deeper tissues pathologies with minimum healthy cells destruction. Recently, Shen et al. reported that UCNPs from highly doping lanthanide ions in NaYF4 nanocrystals can be optically manipulated and demonstrate much higher optical trap stiffness compared to gold nanoparticles [66]. Moreover, theoretical simulations of the electric field distribution in the optical trapped UCNPs of different shapes e.g nanospheres, nanorods were assessed considering NIR laser irradiation [67]. The photoluminescence of UPCNPs could provide new fascinating theranostic interventions by single-cell manipulation and sensing.
(b) During the last few years, optical tweezers were proposed as a tool to probe the Casimir interactions between microspheres inside a liquid medium, membrane proteins and erythrocytes [68-70]. The Casimir effect is a quantum phenomenon arising from quantum fluctuations which can give rise to long-range attractive forces between two uncharged particles [71]. New findings suggest that Casimir forces are responsible for the rouleaux formation of the erythrocytes (red blood cells - RBCs) balancing the electrostatic repulsion between the negative charged RBCs. It is well known that, under certain conditions, erythrocytes form aggregates that look like cylindrical stacks of cell discoids (like coins), the so called "rouleaux" [72]. Additionally, there are many indications that RBC aggregation could play a role in thrombus formation. The cylindrical stack of negatively charged red blood cells could own to the Casimir effect [70].

Recent studies demonstrate that the Casimir forces between two particles can be measured by the optical trapping of the particles in suspensions [73]. Optical tweezers allow pN force measurements. Examples include the force exerted by a single DNA molecule and forces in kinesin or other moto proteins [74]. Physicists manifest that the proteins of cellular membranes can communicate with each other by using Casimir forces. Although they performed experimentally fN force measurements in a simpler scenario than that corresponding to cell biology experiments, they concluded that optical tweezers technique has the potential for novel quantitative applications in cell and molecular biology. Figure 5 illustrates all the interconnected multidisciplinary fields of the advanced Biophotonics techniques for cellular and molecular manipulation through optical tweezers.

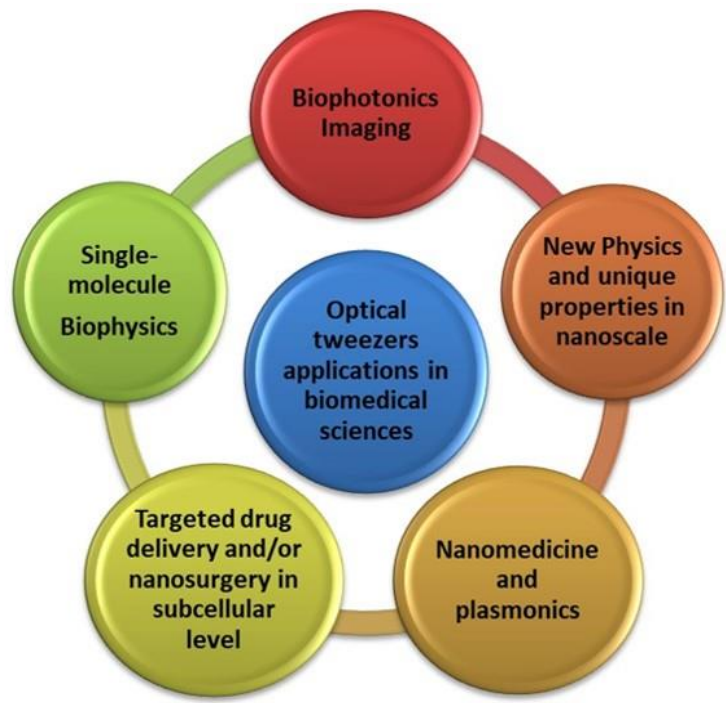

Figure 5. Schematic illustration of some interconnected components of advanced Biophotonics techniques for cellular and molecular manipulation through optical tweezers.

**Acknowledgment**

This research has been co-financed by the European Regional Development Fund of the European Union and Greek national funds through the Operational Program Competitiveness, Entrepreneurship and Innovation, under the call RESEARCH – CREATE –INNOVATE (project code: T1EDK-01223).

arxiv: 2106.05562